\documentclass[aps,prd,groupedaddress,showpacs,nofootinbib,twocolumn]{revtex4}
\usepackage{graphicx,amssymb}

\begin{document}

\title{Gravitational lensing in fourth order gravity}

\author{S. Capozziello$^1$, V.F. Cardone$^2$}
\thanks{Corresponding author\,: {\tt winny@na.infn.it}}

\author{A. Troisi$^1$}

\affiliation{$^1$Dipartimento di Scienze Fisiche, Univ. di Napoli "Federico II", and INFN, Sez. di Napoli,
Compl. Univ. di Monte S. Angelo, Ed. N, via Cinthia, 80121 - Napoli, Italy \\
$^2$Dipartimento di Fisica "E.R. Caianiello",  Univ. di Salerno, and INFN, Sez. di Napoli,
Gruppo Coll. di Salerno, via S. Allende, 84081 - Baronissi (Salerno), Italy}

\begin{abstract}

Gravitational lensing is investigated in the weak field limit of
fourth order gravity in which the Lagrangian of the gravitational
field is modified by replacing the Ricci scalar curvature $R$ with
an analytical expression $f(R)$. Considering the case of a
pointlike lens, we study the behaviour of the deflection angle in
the case of power\,-\,law Lagrangians, i.e. with $f(R) \propto
R^n$. In order to investigate possible detectable signatures, the
position of the Einstein ring and the solutions of the lens
equation are evaluated considering the change with respect to the
standard case. Effects on the amplification of the images and the
Paczynski curve in microlensing experiments are also estimated.

\end{abstract}


\maketitle

\section{Introduction}

The last decade observational results, including Type Ia
Supernovae \cite{SNeIa}, cosmic microwave background anisotropy
spectrum \cite{CMBR,WMAP} and large scale structure \cite{LSS},
have radically changed our view of the universe leading to revise
the old cosmological standard model. On the contrary, the nowadays
standard scenario, referred to as the {\it concordance model},
although still spatially flat, assumes that the visible baryonic
matter constitutes only $\sim 4\%$ of the matter\,-\,energy
content, while the most of the budget is in the form of {\it dark}
components. This dark sector is assumed to be constituted for one
third by {\it dark matter}, characterized by ordinary
thermodynamical properties and mainly clustered in galaxies and
clusters of galaxies, and for two third by a negative pressure
fluid dubbed {\it dark energy}, which drives the large scale
cosmic acceleration and does not cluster at large and small scales
hence not influencing galactic dynamics.

Investigating the nature and the properties of the {\it dark side
of the universe} represents the most daunting and yet fascinating
challenge of modern cosmology so that it is not surprising that
there are a lot of theoretical proposals on the ground.
Nevertheless, both in the case of the old problem of dark matter
and the more recent issue of dark energy, few approaches seem to
furnish a quite natural and coherent framework within which to
explain the observational results. Still more rare are models
which provide a unified view of both dark matter and dark energy
as a two separate manifestations of a single scale\,-\,dependent
gravitational phenomenon.

From this point of view, higher order theories of gravity (both in
the metric \cite{capozcurv,review,sante,MetricRn} and the Palatini
\cite{PalRn,lnR,Allemandi} formulations) represent an interesting
approach able to fruitfully cope with both dark matter and dark
energy problems\footnote{An alternative yet similar approach is
the one based on scalar\,-\,tensor theories of gravity (see, e.g.,
\cite{faraoni} and references therein).}. On one hand, it has been
widely demonstrated that such theories can agree with the
cosmological observations of the Hubble flow
\cite{noiijmpd,noifranca} and the large scale structure evolution
\cite{frlss}. On the other hand, in the weak field limit, the
gravitational potential turns out to be modified
\cite{stelle,schmidt,mannheim,noipla,noimnras} in such a way that
interesting consequences on galactic dynamics may be achieved
without violating, at the same time, the constraints on the PPN
parameters \cite{frppn} from Solar System tests.

It is worth remembering that one of the first experimental
confirmations of Einteinian general relativity was the deflection
of light observed during the Solar eclipse of 1919. Since then,
gravitational lensing (i.e., the deflection of light rays crossing
the gravitational field of a compact object referred to as {\it
lens}) has become one of the most astonishing successes of general
relativity and it represents nowadays a powerful tool able to put
constraints on different scales, from stars to galaxies and
cluster of galaxies, to the large scale structure of the universe
and cosmological parameters (see \cite{SEF} and \cite{petters} for
comprehensive textbooks).

Being intimately related to the underlying theory of gravity in
its Einstein formulation, it is quite obvious that modifying the
Lagrangian of the gravitational field also affects the theory of
gravitational lensing. It is therefore mandatory to investigate
how gravitational lensing works in the framework of higher order
theories of gravity. On the one hand, one has to verify that the
phenomenology of gravitational lensing is preserved in order to
not contradict those observational results that do agree with the
predictions of the {\it standard}\footnote{In the remaining of the
paper, we will refer to the theory of gravitational lensing based
on Einstein general relativity as the {\it standard} or {\it
Einsteinian} theory, while as {\it corrected} theory we mean the
one based on fourth order gravity.} theory of lensing. On the
other hand, it is worth exploring whether deviations from
classical results of the main lensing quantities could be somewhat
detected and work as clear signatures of a modified theory of
gravity. As a first step towards such an ambitious task, we
consider here power\,-\,law fourth order theories, i.e. we replace
the Ricci scalar $R$ in the gravity Lagrangian with the function
$f(R) \propto R^n$, and investigate how this affects the
gravitational lensing in the case of a pointlike lens.

The paper is organized as follows. After a short summary of $f(R)$
theories, we derive, in Sect.\,III, the general expression for the
deflection angle of a pointlike lens in the weak field regime. In
particular, we discuss why this result holds whatever is the
theory of gravity which enters in determining the expression for
the gravitational field. Sect.\,IV is devoted to evaluating the
deflection angle and deriving the lens equation for the case of
the $f(R)$ theory we are considering. Since our results refer to
the case of a pointlike lens, we may compare and contrast them
with the classical ones in the microlensing regime. This issue is
extensively discussed in Sect.\,V where we study the deviations
from the classical results regarding the position of the images,
the Einstein angle, the total amplification and the Paczynski
lightcurve, while, in Sect.\,VI, we summarize and conclude.

\section{Basics of $f(R)$ theories}

Higher order theories of gravity, also referred to as $f(R)$
theories, represent the most natural generalization of the
Einsteinian general relativity. To this aim, one considers the
gravity action\,:

\begin{equation}
\label{f(R)action} {\cal{A}} = \int{d^4x \sqrt{-g} \left [ f(R) +
{\cal{L}}_{m} \right ]}
\end{equation}
where $f(R)$ is a generic function of the Ricci scalar curvature
$R$ differentiable at least up to the second order and
${\cal{L}}_m$ is the standard matter Lagrangian. The choice $f(R)
= R + 2\Lambda$ gives the general relativity including the
contribution of the cosmological constant $\Lambda$. Varying the
action with respect to the metric components $g_{\mu \nu}$, one
gets the generalized Einstein equations that can be more
expressively recast as \cite{capozcurv}\,:

\begin{eqnarray}
G_{\mu \nu} & = & \displaystyle{\frac{1}{f'(R)}}
\displaystyle{\Bigg \{ \frac{1}{2} g_{\mu \nu} \left [ f(R) - R
f'(R) \right ] + f'(R)_{; \mu \nu}} \nonumber \\ ~ & - &
\displaystyle{g_{\mu \nu} \Box{f'(R)} \Bigg \}} +
\displaystyle{\frac{T^{(m)}_{\mu \nu}}{f'(R)}} \label{eq:f-var2}
\end{eqnarray}
where $G_{\mu\nu} = R_{\mu \nu} - (R/2) g_{\mu \nu}$ and
$T^{(m)}_{\mu \nu}$ are the Einstein tensor and the standard
matter stress\,-\,energy tensor respectively and the prime denotes
derivative with respect to $R$. The two terms ${f'(R)}_{; \mu
\nu}$ and $\Box{f'(R)}$ imply fourth order derivatives of the
metric $g_{\mu \nu}$ so that these models are also referred to as
{\it fourth order gravity}. Starting from Eq.(\ref{eq:f-var2}) and
adopting the Robertson\,-\,Walker metric, it is possible to show
that the Friedmann equations may still be written in the usual
form provided that an{\it effective curvature fluid} is added to
the matter term with energy density $\rho_{curv}$ and pressure
$p_{curv} = w_{curv}\rho_{curv}$ depending on the choice of $f(R)$
as\,:

\begin{equation}
\rho_{curv} = \frac{1}{f'(R)} \left \{ \frac{1}{2} \left [ f(R)  -
R f'(R) \right ] - 3 H \dot{R} f''(R) \right \} \ , \label{eq:
rhocurv}
\end{equation}
\begin{equation}
w_{curv} = -1 + \frac{\ddot{R} f''(R) + \dot{R} \left [ \dot{R}
f'''(R) - H f''(R) \right ]} {\left [ f(R) - R f'(R) \right ]/2 -
3 H \dot{R} f''(R)} \ . \label{eq: wcurv}
\end{equation}
As a particular case, we consider power\,-\,law $f(R)$ theories,
i.e. we set\,:

\begin{equation}
f(R) = f_0 R^n
\label{eq: frn}
\end{equation}
with $n$ the slope of the gravity Lagrangian ($n = 1$ being the
Einstein theory) and $f_0$ a constant with the dimensions chosen
in such a way to give $f(R)$ the right physical dimensions. It has
been shown that such a choice leads to matter only models able to
fit well the Hubble diagram of Type Ia Supernovae without the need
of dark energy \cite{noiijmpd} and could also be reconciled with
the constraints on the PPN parameters \cite{frppn}.

\section{The deflection angle}

Although generalizing the gravity Lagrangian has a deep impact on
both the cosmology and the local dynamics, it is nevertheless easy
to understand that the derivation of the lens equation is formally
the same as that in standard general relativity. Indeed, the basic
assumption in deriving the lens equation is that the gravitational
field is weak and stationary. In this case, the spacetime metric
reads\,:

\begin{equation}
ds^2 = \left ( 1 - \frac{2 \Phi}{c^2} \right ) c^2 dt^2 - \left (
1 + \frac{2 \Phi}{c^2} \right ) \delta_{ij} dx^i dx^j \label{eq:
weak}
\end{equation}
where $\Phi$ is the gravitational potential and, as usual, we have
neglected the gravitomagnetic term\footnote{Gravitomagnetic
effects originate from the mass current in the lens and, under
some circumstances, could give rise to effects that could be
detected with future interferometric satellites (see, e.g.,
\cite{gravmag} and references therein).}. Since light rays move
along the geodetics of the metric (\ref{eq: weak}), the lens
equation may be simply derived by solving $ds^2 = 0$ and keeping
only terms up to second order in $v/c$. It is worth stressing that
such a derivation holds whatever is the theory of gravity provided
that one can still write Eq.(\ref{eq: weak}) in the approximation
of weak and stationary fields. As a fundamental consequence, the
lensing deflection angle will be given by the same formal
expression that holds in general relativity \cite{SEF,petters}\,

\begin{equation}
\vec{\alpha} = \frac{2}{c^2} \int{\vec{\nabla}_{\perp} \Phi dl}
\label{eq: vecalpha}
\end{equation}
with\,:

\begin{equation}
\vec{\nabla}_{\perp} \equiv \vec{\nabla} - \hat{e} (\hat{e} {\cdot} \vec{\nabla})
\label{eq: nablaperp}
\end{equation}
where $\hat{e}$ is the spatial vector\footnote{Here, quantities
with an over\,-\,arrow are vectors, while the versor will be
denoted by an over\,-\,hat.} tangent to the direction of the light
ray and $dl = \sqrt{\delta_{ij} dx^i dx^j}$ is the Euclidean line
element. The integral in Eq.(\ref{eq: vecalpha}) should be
performed along the light ray trajectory which is {\it a priori}
unknown. However, for weak gravitational fields and small
deflection angles, one may integrate along the unperturbed
direction. In this case, we may set the position along the light
ray as\,:

\begin{equation}
\vec{r} = \vec{\xi} + l \hat{e}
\label{eq: defxi}
\end{equation}
with $\vec{\xi}$ orthogonal to the light ray. Assuming the
potential only depending on $r = | \vec{r} | = (\xi^2 +
l^2)^{1/2}$ (as for a pointlike or a spherically symmetric lens),
the deflection angle reduces to\,:

\begin{equation}
\vec{\alpha} = \frac{2 \vec{\xi}}{c^2} \int_{-\infty}^{\infty}{\left ( \frac{1}{r} \frac{d\Phi}{dr} \right ) dl}
\label{eq: alphacent}
\end{equation}
where we have assumed that the geometric optics approximation
holds, the light rays are paraxial and propagate from infinite
distance.

Eq.(\ref{eq: alphacent}) is general and holds whatever is the
underlying theory of gravity. This nice result does not mean that
the deflection angle is the same as in the Einsteinian general
relativity. Indeed, the link with the underlying gravity theory is
represented by the gravitational potential $\Phi$ which is
determined by solving the corresponding generalized Einstein
equations for the weak field metric (\ref{eq: weak}). As a result,
$\Phi$ can be no more the standard Newtonian potential and
deviations from the Keplerian scaling $1/r$ enter the game leading
to interesting consequences. From the lensing point of view, this
means that, although Eq.(\ref{eq: alphacent}) still holds, the
deflection angle differs from the one of general relativity in a
way that critically depends on the expression chosen for the
function $f(R)$ entering the gravity Lagrangian.

Before deriving explicitly the deflection angle in $f(R)$
theories, it is worth discussing an important point. The above
considerations rely on the assumption that the weak field metric
may be still be written as in Eq.(\ref{eq: weak}) in the case of
alternative theories of gravity. That is, one is implicitly
assuming that the solution of the field equations in the low
energy limit shares the same formal structure as the Schwarzschild
solution which Eq.(\ref{eq: weak}) reduces to when $\Phi(r)$ is
the Newtonian $1/r$ potential. Actually, under the hypothesis of
weak gravitational fields and slow motions (both well verified in
the astrophysical phenomena of interest), by virtue of the
Birkhoff\,-\,Jensen theorem, the weak field metric is\,:

\begin{equation}
ds^2 = A(r) dt^2 - B(r) dr^2 - r^2 d\Omega^2 \label{eq: schwartz}
\end{equation}
where $d\Omega^2 = d\theta^2 + \sin^2{\theta} d\varphi^2$ is the
line element on the unit sphere. Inserting Eq.(\ref{eq: schwartz})
into the Einstein field equations one obtains the classical
Schwarzschild solution by setting $A(r) = 1/B(r) = 1 +
2\Phi(r)/c^2$ and using the limit $\Phi/c^2 << 1$ with $\Phi(r)$
the Newtonian potential. One could argue that this is no more the
case when alternative theories of gravity are considered so that
one should take $A(r) \ne 1/B(r)$ and then deriving the lens
equation in the general PPN formalism as in \cite{kp}. However,
considering higher order theories of gravity leads to higher order
field equations and hence widens the set of solutions. It is
therefore possible, {\it a priori}, to find out solutions of the
low energy limit of the field equations under the assumption $A(r)
= 1/B(r)$ so that the spacetime metric indeed reads as in
Eq.(\ref{eq: weak}) and the deflection angle is given by
Eq.(\ref{eq: alphacent}) also for $f(R)$ theories.

\section{The pointlike lens}

Eq.(\ref{eq: alphacent}) allows to evaluate the deflection angle
provided that the source mass distribution and the theory of
gravity have been assigned so that one may determine the
gravitational potential. As a first step, we consider here the
case of the pointlike lens. Note that, although being the simplest
one, the pointlike lens is the standard approximation for stellar
lenses in microlensing applications \cite{SEF,MR02}. Moreover,
since in the weak field limit $\vec{\alpha}$ is an additive
quantity, the deflection angle for an extended lens may be
computed integrating the pointlike result weighted by the
deflector surface mass distribution under the approximation of
thin lens \cite{SEF} (i.e., the mass distribution extends over a
scale which is far smaller than the distances between observer,
lens and source).

Given the symmetry of the problem, it is clear that we may deal
with the magnitude of the deflection angle and of the other
quantities of interest rather than with vectors. In the
approximation of small deflection angles, simple geometrical
considerations allow to write the lens equation as\,:

\begin{equation}
\theta - \theta_s = \frac{D_{ls}}{D_s} \alpha
\label{eq: lenseq}
\end{equation}
which gives the position $\theta$ in the lens plane of the images
of the source situated at the position $\theta_s$ on the source
plane\footnote{Both $\theta$ and $\theta_s$ are measured in
angular units and  could be redefined as $\theta = \xi/D_l$ and
$\theta_s = \eta/D_s$ with $\xi$ and $\eta$ in linear units.}.
Note that the lens and the source planes are defined as the planes
orthogonal to the optical axis which is the line joining the
observer and the centre of the lens. Here and in the following,
$D_l$, $D_s$, $D_{ls}$ are the angular diameter distances between
observer\,-\,lens, observer\,-\,source and lens\,-\,source
respectively.

In order to evaluate the deflection angle, we need an explicit
expression for the gravitational potential $\Phi$ generated by a
pointlike mass $m$. This has been derived in detail in
\cite{noimnras} for the case of power\,-\,law $f(R)$ theories.
Solving the low energy limit of the field equations
(\ref{eq:f-var2}) in vacuum ($T^{(m)}_{\mu \nu} = 0$) for $f(R)$
given by Eq.(\ref{eq: frn}), one obtains for the gravitational
potential\,:

\begin{equation}
\Phi(r) = - \frac{G m}{2 r} \left [ 1 + \left ( \frac{r}{r_c}
\right )^{\beta} \right ] \label{eq: phi}
\end{equation}
with\,:

\begin{equation}
\beta = \frac{12n^2 - 7n - 1 - \sqrt{36n^4 + 12n^3 - 83n^2 + 50n +
1}}{6n^2 - 4n + 2} \ . \label{eq: bnfinal}
\end{equation}
For $n = 1$, $\beta = 0$ and the potential reduces to the
Newtonian one as expected. While the slope $\beta$ of the
correction term is a universal quantity (since it depends on the
exponent $n$ entering the gravity Lagrangian), the scalelength
$r_c$ is related to one of the integration constant that have to
be set to solve the fourth order differential equations of the
theory. As such, $r_c$ is expected to be related to the
peculiarities of each system and can therefore take different
values depending on its mass and typical scale.

 Inserting Eq.(\ref{eq: phi}) into
(\ref{eq: alphacent}) and using $\xi = D_l \theta$, after some
algebra, we finally get for the deflection angle\,:

\begin{equation}
\alpha = \frac{2 G m}{c^2 r_c} \left ( \frac{\xi}{r_c} \right
)^{-1} \left [ 1 + \frac{\sqrt{\pi} (1 - \beta) \Gamma(1 -
\beta/2)}{2 \Gamma(3/2 - \beta/2)} \left ( \frac{\xi}{r_c} \right
)^{\beta} \right ] \label{eq: alphapoint}
\end{equation}
which is defined\footnote{For $\beta$ outside this range, the
deflection angle may still be computed, but there is not an
analytical expression.} only for $0 \le \beta \le 2$. In the
following, however, we will only consider the narrowest range $0
\le \beta \le 1$ since, as discussed in \cite{noimnras}, in this
case, the modified potential offers the possibility to fit
galactic rotation curves without dark matter. As a cross check,
note that, for $n = 1$, it is $\beta = 0$ and\,:

\begin{displaymath}
\lim_{\beta \rightarrow 0}{\alpha(\xi, \beta)} = \alpha_{GR} =
\frac{4 G m}{c^2 \xi}
\end{displaymath}
so that the classical\footnote{Hereafter, we will denote with the
subscript {\it GR} quantities evaluated for $n = 1$, while we use
the subscript {\it FOG} for the correction due to the fourth order
theory. With this notation, for instance, the total deflection
angle is $\alpha = \alpha_{GR} + \alpha_{FOG}$. Moreover, we refer
to the results obtained for $n = 1$ as {\it classical}.} result of
Einsteinian relativity is recovered. On the contrary, for $\beta =
1$, we get $\alpha = \alpha_{GR}/2$ so that the net effect of the
correction term is to decrease the deflection angle with respect
to the classical one.

As we have quoted above and explicitly shown in \cite{noimnras},
the modified gravitational potential leads to an increase of
galactic rotation curve acting as a sort of {\it effective} dark
matter filling the gap between the Newtonian and the observed
rotation curve. Given that the classical deflection angle is
proportional to the mass of the system, one could expect that the
modified potential leads to an increase of the deflection angle
while we get the opposite result. However, there is an important
difference explaining this counterintuitive behaviour. Indeed, the
deflection angle depends on the integral of $ {\cal{I}} = (1/r)
d\Phi/dr$. Comparing the Newtonian potential with the modified
one, we get\,:

\begin{displaymath}
\frac{\Phi(r)}{\Phi_{GR}(r)} = \frac{1}{2} \left [ 1 + \left (
\frac{r}{r_c} \right )^{\beta} \right ] \ ,
\end{displaymath}

\begin{displaymath}
\frac{{\cal{I}}(r)}{{\cal{I}}_{GR}(r)} = \frac{1}{2} \left [ 1 +
(1 - \beta) \left ( \frac{r}{r_c} \right )^{\beta} \right ] \ ,
\end{displaymath}
so that

\begin{displaymath}
\Phi(r) > \Phi_{GR}(r) \iff r > r_c \ ,
\end{displaymath}

\begin{displaymath}
{\cal{I}}(r) > {\cal{I}}_{GR}(r) \iff r > (1 - \beta)^{-\beta} r_c
\ .
\end{displaymath}
In order an increase of the rotation curve (and hence a good fit
without dark matter), the first condition should be met which is
easy to realize on galactic scales given typical values of $r_c$.
On the contrary, for a pointmass lens, since $\beta < 1$, the
integral entering the deflection angle ${\cal{I}}(r)$ is actually
always smaller than the corresponding classical one
${\cal{I}}_{GR}(r)$ since $(1 - \beta)^{-\beta} r_c$ is very
large. As a consequence, then, the corrected deflection angle
turns out to be lower than the classical one.

Inserting Eq.(\ref{eq: alphapoint}) into Eq.(\ref{eq: lenseq}),
the lens equation may be conveniently written as\,:

\begin{equation}
\left [ 1 -  {\cal{N}}(\beta, \vartheta_c) \left (
\frac{\vartheta}{\vartheta_c} \right )^{\beta - 2} \right ]
\vartheta^2 - \vartheta_s \vartheta - \frac{1}{2} = 0 \label{eq:
lenseqbis}
\end{equation}
with

\begin{equation}
{\cal{N}}(\beta, \vartheta_c) =  \frac{\sqrt{\pi} (1 - \beta)
\Gamma(1 - \beta/2)}{4 \vartheta_c^2 \Gamma(3/2 - \beta/2)}
\label{eq: defenne}
\end{equation}
having defined $\vartheta = \theta/\theta_{E,GR}$, with
$\theta_{E,GR}$ the Einstein angle in the general relativity case
given by\,:

\begin{equation}
\theta_{E,GR} = \sqrt{\frac{4 G m D_{ls}}{c^2 D_l D_s}} \ .
\label{eq: einst}
\end{equation}
Note that, for $n = 1$ ($\beta = 0$), the corrected lens equation
(\ref{eq: lenseqbis}) reduces to the classical one\,:

\begin{displaymath}
\vartheta^2 - \vartheta_s \vartheta - 1 = 0
\end{displaymath}
thus ensuring the self consistency of the theory. As a general
remark, the modified lens equation differs from the classical one
because of the second term in the square brackets which furnishes
a correction depending on both $\beta$ and $r_c$ (through
$\vartheta_c = r_c/D_l \theta_{E,GR}$). Actually, the magnitude of
the correction strongly depends not only on the scaling radius
$r_c$, but also on the lens mass and the distances $(D_l, D_s,
D_{ls})$ since they all enter $\vartheta_c$.

\section{Microlensing observables}

The pointlike lens equation (\ref{eq: lenseqbis}) differs from the
standard general relativistic one for the second term in the
square parenthesis. Should this term be negligible, all the usual
results of gravitational lensing are recovered. It is therefore
interesting to investigate in detail how the corrective term
affects the estimate of observable related quantities since,
should they come out to be detectable, they could represent a
signature of $R^n$ gravity.

Since we are considering the pointlike lens, a typical lensing
system is represented by a compact object (both visible or not) in
the Galaxy halo acting as a lens on the light rays coming from a
stellar source in an external galaxy, like the Magellanic Clouds
(LMC and SMC) or Andromeda. It is easy to show that, in such a
configuration, the standard Einstein angle $\theta_{E,GR}$ and the
images angular separation are of the order of few ${\times}
10^{-5} \ arcsec$ so that we are in the regime known as {\it
microlensing} \cite{MR02}.

In the following, we will consider different observable quantities
in microlensing applications in order to see whether it is
possible to detect deviations from the standard theory of
gravitational lensing. As a fiducial system, we take a stellar
lens in the galactic halo (with $m = 1 \ {\rm M_{\odot}}$ and $D_l
= 20 \ {\rm kpc}$) and a source star in the LMC (with $D_s = 50 \
{\rm kpc})$. The main results may be easily scaled to other values
of $(m, D_l, D_s)$ by the corresponding Newtonian Einstein angle.

The dependence on the theory is parametrized by the two parameters
$(n, r_c)$ entering the modified gravitational potential (\ref{eq:
phi}). Actually, we prefer to use $\beta$ rather than $n$ as first
parameter since its value is limited in the narrow range $(0, 1)$
with $\beta = 0$ corresponding to the General Relativity case $n =
1$. In order to estimate a reasonable range for $r_c$, we proceed
as follows. Let us rewrite the potential as\,:

\begin{displaymath}
\Phi = - \frac{G m}{2 r_c} \eta^{-1} \left ( 1 + \eta^{\beta}
\right ) = - \frac{\Phi_1}{2} \eta^{-1} \left ( 1 + \eta^{\beta}
\right )
\end{displaymath}
with $\eta = r/r_c$ and $\Phi_1 = G m/r_c$. Since $\Phi_1$ has the
dimension of a squared velocity, we introduce a further parameter
$v_1 = \Phi_1^{1/2}$ so that the potential conveniently reads

\begin{displaymath}
\Phi = - \frac{c^2}{2} \left ( \frac{v_1}{c} \right )^2 \eta^{-1}
\left ( 1 + \eta^{\beta} \right ) \ .
\end{displaymath}
Comparing the two expressions for the potential, we get\,:

\begin{equation}
r_c = \frac{G m}{c^2 \tilde{v}_1^2}
\label{eq: rcv1}
\end{equation}
with $\tilde{v}_1 = v_1/c$. Rather than $r_c$, we will use in the
following $\log{\tilde{v}_1}$ as parameter since its range is
easier to evaluate. Let us consider, for instance, the Sun as the
field source and a planet as a test particle. Since the Earth
velocity is $\simeq 30 \ {\rm km/s}$, we can indeed take $(-6,
-4)$ as a reasonable range for $\log{\tilde{v}_1}$. As a final
remark, note that Eq.(\ref{eq: rcv1}) clearly shows why $r_c$
depends on both the mass and the scale of the system under
examination.

\subsection{The Einstein angle and the position of the images}

As a first application, let us consider the solution of the lens
equation starting from the case of perfect alignment among
observer, lens and source. Setting $\vartheta_s = 0$, Eq.(\ref{eq:
lenseqbis}) reduces to\,:

\begin{figure}
\centering \resizebox{8.5cm}{!}{\includegraphics{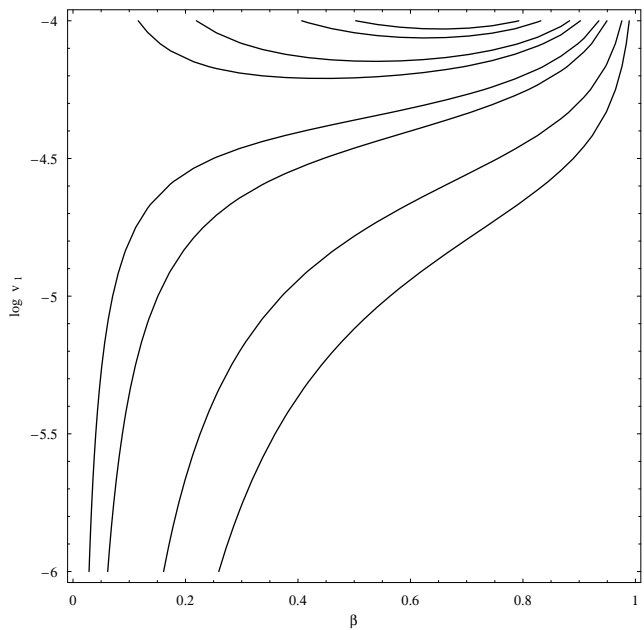}}
\caption{Contours of equal $\varepsilon_E$ in the plane $(\beta,
\log{\tilde{v}_1})$ for $\varepsilon_E = (-0.25, -0.20, -0.10,
-0.05, 0.05, 0.10, 0.20, 0.25)$ from the bottom curve to the top
one.} \label{fig: teconts}
\end{figure}

\begin{equation}
\left [ 1 - {\cal{N}}(\beta, \vartheta_c) \left (
\frac{\vartheta}{\vartheta_c} \right )^{\beta - 2} \right ]
\vartheta^2 = \frac{1}{2}
\label{eq: eqeinst}
\end{equation}
which, for $\beta = 0$ ($n = 1$) reduces to\,:

\begin{displaymath}
\vartheta^2 = 1 \iff \theta = \theta_{E,GR}
\end{displaymath}
i.e. the image is the well known Einstein ring with angular radius
equal to the Einstein angle defined in Eq.(\ref{eq: einst}). In
the general case, Eq.(\ref{eq: eqeinst}) must be solved
numerically for given values of the model parameters $(\beta,
\log{\tilde{v}_1})$. As a general result, the Einstein ring still
forms, but the corrected Einstein angle $\theta_E$ deviates from
the classical one, i.e. $\vartheta_E \ne 1$. It is therefore
interesting to study the magnitude of such deviations. To this
aim, we define\,:

\begin{displaymath}
\varepsilon_E = \frac{\theta_E - \theta_{E,GR}}{\theta_{E,GR}} =
\vartheta_E - 1
\end{displaymath}
which quantifies the relative deviation with respect to the
classical result. Contours of equal $\varepsilon_E$ in the plane
$(\beta, \log{\tilde{v}_1})$ are shown in Fig. \ref{fig: teconts}.

Some interesting remarks may be drawn from this plot. As a general
result, $\varepsilon_E$ could take both positive and negative
values, i.e. $R^n$ gravity may increase or decrease the Einstein
angle with respect to the classical result. Actually, over almost
the full parameter space, $\varepsilon_E$ is negative and, in
particular, taking $(\beta, \log{\tilde{v}_1}) \simeq (0.58, -5)$
as fiducial values\footnote{Indeed, fitting the LSB rotation
curves, one gets $\beta = 0.58 \pm 0.15$ \cite{noimnras}, while
$\log{\tilde{v}_1} \simeq -5$ is suggested by considering a
Jovian\,-\,like planet as test particle in the potential of a
Sun\,-\,like lens.}, we find $\varepsilon_E \simeq -0.25$ which
leaves open the possibility to detect the corrections due to $R^n$
gravity.

Before discussing this issue, it is worth examining how
$\varepsilon_E$ depends on the model parameters. For fixed values
of $\beta$, the relative deviation is clearly an increasing
function of $\log{\tilde{v}_1}$, a result that can be easily
explained qualitatively. Indeed, the larger is
$\log{\tilde{v}_1}$, the smaller is $r_c$ and hence the earlier
the correction term into the modified potential starts
contributing thus explaining why the Einstein angle deviates more
and more from the classical one as $\log{\tilde{v}_1}$ increase. A
similar discussion may be done to explain why $\varepsilon_E$
increases (in absolute value) with $\beta$ for fixed
$\log{\tilde{v}_1}$. To this aim, let us remember that the
deflection angle $\alpha$ changes from $\alpha_{GR}$ for $\beta =
0$ to $\alpha_{GR}/2$ for $\beta = 1$ so that the lensing {\it
strength} decreases with $\beta$. As a consequence, the Einstein
angle becomes smaller as $\beta$ gets higher and hence
$\varepsilon_E$ takes more negative values.

Modifying the Einstein angle has a deep impact on the microlensing
phenomenology. As a first interesting application, let us remember
that the duration of a microlensing event is estimated as $t_E =
R_E/v_{\perp}$ with $R_E = D_l \theta_E$ the Einstein radius and
$v_{\perp}$ the transverse velocity of the lens with respect to
the line of sight. Supposing that the distances $(D_l, D_s)$ and
the transverse velocity $v_{\perp}$ were known (for instance, from
deviations with respect to the standard lightcurve), a measurement
of the Einstein time $t_{E,obs}$ translates into an estimate of
the mass $m$ of the lens. Denoting with $\mu$ the true mass (in
units of $1 \ {\rm M_{\odot}}$) and with $\mu_{GR}$ the one
estimated from using incorrectly the classical expression for the
Einstein angle, it is respectively\,:

\begin{displaymath}
t_{E,obs} = \mu^{1/2} t_E(m = 1 \ {\rm M_{\odot}}) =
\mu_{GR}^{1/2} t_{E,GR}(m = 1 \ {\rm M_{\odot}})
\end{displaymath}
so that we get\,:

\begin{equation}
\mu = \frac{\mu_{GR}}{\vartheta_E^2(m = 1 \ {\rm M_{\odot}})} \ .
\end{equation}
As can be inferred from Fig. \ref{fig: teconts}, $\vartheta_E^2(m
= 1 \ {\rm M_{\odot}}) < 1$ over almost the full parameter space
$(\beta, \log{\tilde{v}_1})$ so that the true lens mass is higher
than the one estimated by the classical method. However, in most
of the applications, one may statistically infer only the lens
transverse velocity, but not its distance so that what is indeed
underestimated by the classical method is the quantity $\mu x (1 -
x)$ with $ x = D_l/D_s$. As such, therefore, taking into account
the correction to the Einstein angle may lead to an increase of
either the estimated lens mass or of the distance ratio $x$.
Disentangling these two effects is quite difficult and need for
non standard microlensing events. A possible way out of this
problem could be resorting to non standard microlensing events,
but, in such a case, the duration of the event may differ from
$t_E$ so that the above qualitative argument does not apply
anymore.

A similar discussion can be made for the optical depth $\tau$ that
measures the probability of finding a lens in the microlensing
tube, a cylinder with main axis along the line of sight and with
radius equal to the Einstein radius. Since the corrected Einstein
radius is smaller than the classical one, the microlensing tube is
narrower and hence the optical depth, for a given spatial
distribution of the lenses, is lower than the classical one.

As a second application, let us study how the images position
change. In the Einsteinian case, the lens equation may be solved
analytically and one gets two images with positions given by\,:

\begin{equation}
\vartheta_{{\pm,GR}} = \frac{1}{2} \left (\vartheta_s {\pm}
\sqrt{\vartheta_s^2 + 4} \right ) \ . \label{eq: poseinst}
\end{equation}
In the current case, the lens equation (\ref{eq: lenseqbis}) must
be solved numerically for given values of the source position
$\vartheta_s$ and the $f(R)$ parameters $(\beta,
\log{\tilde{v}_1})$. We still get two images on opposite sides of
the lens with one image lying inside and the other one outside the
Einstein ring. The geometric configuration is therefore the same
as in the standard case, but the positions are slightly changed.
To quantify this effect, we have studied the contours of equal
$\varepsilon_{\pm}$ in the plane $(\beta, \log{\tilde{v}_1})$ with
$\varepsilon_{\pm}$ defined in a similar way as $\varepsilon_E$.
Not surprisingly, we get the same results as for the Einstein
angle, i.e. the contours are parallel to those of $\varepsilon_E$
but shifted by an amount depending on $\vartheta_s$. This is
easily explained noting that the lens equation is quite similar to
the one for the Einstein angle, while the term depending on the
$R^n$ gravity parameters is the same.

\subsection{Images amplification and Paczynski curve}

Gravitational lenses work also as a sort of {\it natural
telescopes} amplifying the luminosity of the source in such a way
to make visible objects that are otherwise too faint to be
detected. As a general rule, the amplification is given by the
inverse of the Jacobian of lens mapping. In the case of a lens
with cylindrical symmetry (and hence also for the pointlike lens),
this reduces to \cite{SEF}\,:

\begin{equation}
{\cal{A}} = \left | \frac{\vartheta_s}{\vartheta}
\frac{d\vartheta_s}{d\vartheta} \right |^{-1} \ . \label{eq:
defamp}
\end{equation}
Starting from Eq.(\ref{eq: lenseqbis}), after some algebra, one gets\,:

\begin{equation}
{\cal{A}} = \left | 1 - a^2 - \left [ (2 - \beta) a + (1 - \beta)
b + \beta \right ] b \right |^{-1} \label{eq: ampcorr}
\end{equation}
with\,:

\begin{equation}
a = (2 \vartheta^2)^{-1} \ ,
\label{eq: defa}
\end{equation}

\begin{equation}
b = {\cal{N}}(\beta, \vartheta_c) \left ( \vartheta/\vartheta_c
\right )^{\beta - 2} \ .
\label{eq: defb}
\end{equation}
Note that we have considered the unsigned amplification since we
are not interested to the parity of the images, but only to the
luminosity variation. As a check of the consistency of the theory,
it is easy to see that\,:

\begin{equation}
\lim_{\beta \rightarrow 0}{{\cal{A}}(\vartheta, \beta)} =
{\cal{A}}_{GR} = \left |1 - \frac{1}{\vartheta^4} \right |^{-1}
\label{eq: ampeinst}
\end{equation}
so that the General Relativity result is recovered for $n = 1$. At
the other extreme, we get\,:

\begin{equation}
\lim_{\beta \rightarrow 1}{{\cal{A}}(\vartheta, \beta)} = \left |1
- \frac{1}{4 \vartheta^4} \right |^{-1}
\end{equation}
so that the amplification is a increasing function of the slope
parameter $\beta$. This is in contrast with the results obtained
for the Einstein angle and the position of the images. This result
is not unexpected since the amplification is defined in terms of
the inverse of the Jacobian mapping and therefore increases when
this latter decreases. Since $\det{J}$ is a decreasing function of
$\beta$, the behaviour of ${\cal{A}}$ is obvious.

As a first important check, we have verified that the corrections
to the amplification do not change the structure of the critical
curves defined as the loci in the lens plane where the
amplification gets formally infinite. Indeed, using Eq.(\ref{eq:
eqeinst}), we see that $b(\vartheta_E) = 1 - a(\vartheta_E)$ that,
inserted into Eq.(\ref{eq: ampcorr}), gives $\det{J}(\vartheta_E)
= 0$ and hence ${\cal{A}}(\vartheta_E) = \infty$. Therefore, the
critical curve is still the Einstein angle as in the classical
case and the only caustic (which is obtained projecting the
critical curve on the source plane) is $\vartheta_s = 0$ as usual.

Since in a microlensing event the two images are merged together
(given the tiny angular separation of the order twice the Einstein
angle), what we observe is the total luminosity which is obtained
by multiplying the source luminosity by the sum of the
amplification of the two images. In the Einsteinian case,
inserting Eqs.(\ref{eq: poseinst}) into the right hand side of
Eq.(\ref{eq: ampeinst}), one gets for the total amplification\,:

\begin{equation}
{\cal{A}}_{tot,GR} = {\cal{A}}_{+} + {\cal{A}}_{-} = \left |
\frac{\vartheta_s^2 + 2}{\vartheta_s \sqrt{\vartheta_s^2 + 4}}
\right | \ . \label{eq: amptoteinst}
\end{equation}
A similar relation for the corrected total amplification cannot be
written since we do not have an analytical expression for the
images position as a function of the source position. Actually,
since the intrinsic source flux is unknown, we are unable to
measure ${\cal{A}}_{tot}$ so that detecting deviations from the
standard theory is not possible at all. Nevertheless, a different
strategy could be implemented. Assuming that the lens crosses the
line of sight moving with a constant velocity, the source position
as function of time $t$ could be written as \cite{Pac86}\,:

\begin{equation}
\vartheta_s = \sqrt{\vartheta_0^2 + \left ( \frac{t - t_0}{t_E} \right )^2}
\label{eq: tsvst}
\end{equation}
with $\vartheta_0 = \vartheta_s(t = t_0)$, $t_0$ the time of
closest approach to the lens to the line of sight and $t_E$ the
Einstein time defined above. As the time goes on, the source
position changes according to Eq.(\ref{eq: tsvst}) and hence the
images position varies. As a consequence, also the total
amplification ${\cal{A}}_{tot}$ becomes a function of time. In the
Einsteinian case, inserting Eq.(\ref{eq: tsvst}) into Eq.(\ref{eq:
amptoteinst}), one gets the well known Paczynski lightcurve
\cite{Pac86} having the peculiar signatures of uniqueness and
symmetry of the bump and achromaticity. Using the numerical
solutions of Eq.(\ref{eq: lenseqbis}), we can work out a corrected
Paczynski lightcurve for the case of power\,-\,law $f(R)$
theories. Some examples are shown in Fig.\,\ref{fig: paczcurve}
where, without loss of generality, we have set $t_0 = 0$ and $t_E
= 10 \ {\rm d}$.

\begin{figure}
\centering \resizebox{8.5cm}{!}{\includegraphics{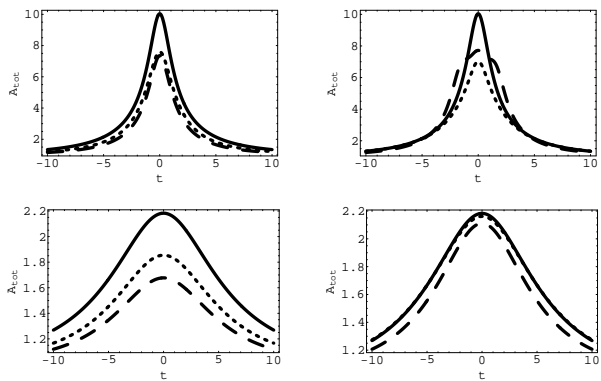}}
\caption{Paczynski lightcurve for Einsteinian gravitational
lensing (solid line) and the $R^n$ gravity with $\beta = 0.25$
(short dashed) and $\beta = 0.75$ (long dashed). Upper (lower)
panels refer to $\vartheta_0 = 0.1$ (0.5), while left (right)
panels are for $\log{\tilde{v}_1} = -5.5$ (-4.5).} \label{fig:
paczcurve}
\end{figure}

As it is apparent from the figure, both the sign and the magnitude
of the deviations from the standard lightcurve depend in a
complicated way on both the theory parameters $(\beta,
\log{\tilde{v}_1})$ and the minimum impact parameter
$\vartheta_0$. Although a general rule cannot be extracted from
the plots, it is interesting to note, however, that, for some
combination of the parameters, the microlensing lightcurve may be
also distorted with respect to the Paczynski one. Nevertheless,
the typical signature of microlensing events (symmetry, uniqueness
of the bump and acromaticity) are preserved.

\begin{figure}
\centering \resizebox{8.5cm}{!}{\includegraphics{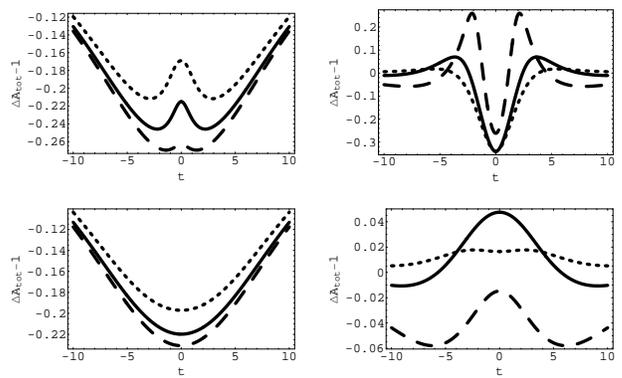}}
\caption{Relative deviation with respect to the Paczynski
lightcurve of the corrected microlensing lightcurve for $R^n$
gravity with $\beta = 0.43$ (short dashed), $\beta = 0.58$ (solid)
and $\beta = 0.73$ (long dashed). $\vartheta_0$ ad
$\log{\tilde{v}_1}$ are set as in Fig. \ref{fig: paczdiff}.}
\label{fig: paczdiff}
\end{figure}

To estimate quantitatively the deviations from the Paczynski
lightcurve, we plot in Fig. \ref{fig: paczdiff} the quantity
$\Delta A_{tot} - 1$ as function of the time $t$ for the same
values of $(t_0, t_E)$. Considering, for instance, solid lines in
the right panels, it is clear that the deviations are of order
$10\%$ which leaves open the possibility for a detection in the
highest magnification events where the measurement errors on the
lightcurve points are sufficiently low. A more detailed
investigation is, however, needed to understand whether such
deviations may be mimicked by other astrophysical effects. Indeed,
we have here considered the case of a point mass acting as a lens
on a point source. Should the lens (or the source) be a double
system, this treatment breaks down and the curve may be distorted
in a predictable way. Comparing these effects with those
introduced by $R^n$ gravity is outside our aim here, but it is a
task worth to be addressed in detail to understand whether there
are detectable signature of power\,-\,law $f(R)$ theories in the
events lightcurves.

\section{Conclusions}

Fourth order gravity has been recently proposed as viable
alternatives to quintessence scalar fields and exotic fluids to
solve the problem of cosmic speed up. In particular, power\,-\,law
$f(R)$ theories have been shown to successfully fit the SNeIa data
without violating the constraints on the PPN parameters. Moreover,
in the weak field limit, they give rise to a modified
gravitational potential that makes it possible to fit the LSB
rotation curves without the need of any dark matter halo.

Having successfully passed this impressive set of observational
tests, the proposed modification to Einstein general relativity
worths to be further investigated considering its effects on
gravitational lensing. As a first step, we have derived an
analytical expression for the deflection angle of a pointlike lens
since this is the basic ingredient for a generalization to the
case of extended systems (such as galaxies and clusters of
galaxies). Because of the deviations of the gravitational
potential from the Newtonian one, the deflection angle turns out
to differ from the Einsteinian one because of an additive
power\,-\,law term depending on the slope $\beta$ and the
scalelenght $r_c$ of the modified potential. For $0 \le \beta \le
1$, the corrected deflection angle is smaller than the classical
one, but the amount of the corrections critically depends not only
on $\beta$ and $r_c$, but also on the lens mass $m$ and the
distances $(D_l, D_s, D_{ls})$. As a consequence of the corrected
deflection angle, the lens equation may no more be solved
analytically and the resort to numerical techniques is needed. The
number of images is still two with an angular separation of the
order of the corrected Einstein angle.

To quantitatively investigate the impact of the corrections on
observable quantities, we have considered the case of a solar mass
object in the Galactic halo acting as a lens on a source star in
the LMC so that we are in the typical conditions of microlensing
regime. Denoting with $\varepsilon_E$ the relative deviation of
the corrected Einstein angle with respect to the classical one, we
find out that both its sign and magnitude depend on the region in
parameter plane $(\beta, \log{\tilde{v}_1})$ considered, but
significant deviations ($\varepsilon_E \sim -20\%$) may be
obtained for the value of $\beta$ ($\simeq 0.58$) suggested by the
fit to the LSB galaxies rotation curves. A similar conclusion also
holds for the deviations from the classical Paczynski lightcurve.
Although a more detailed analysis (also including non standard
microlensing events and observational uncertainties) is needed,
these preliminary results suggest the intriguing possibility to
detect $R^n$ gravity signatures through a careful examination of
galactic microlensing. This may open the way to a completely new
regime in which looking for corrections to the Einsteinian general
relativity that may be complementary to the other cosmological
probes.

It is worth stressing that, since significant deviations from the
standard theory of microlensing are possible, the current
estimates of the optical depth $\tau$ towards LMC (or any other
target such as the galactic bulge, SMC or Andromeda) and of the
mean lens mass inferred from the time duration distribution of
microlensing events could be altered. In particular, the classical
estimates of the optical depth (the mean lens mass) should be
revised downward (upward) for fiducial values of the parameters
$(\beta, \log{\tilde{v}_1}$). Interestingly, microlensing surveys
towards LMC have concluded that less than $20\%$ of the putative
dark halo is made out of compact baryonic objects (see, e.g.,
\cite{jetz05} and references therein) and there are models
explaining the observed optical depth as a result of known stellar
populations only belonging to our Galaxy \cite{inside} and/or to
the LMC \cite{self}. On the other hand, our modified gravity makes
it possible to fit the Milky Way rotation curve without any dark
matter halo. Indeed, one may speculate that self\,-\,lensing
models should be preferred in the context of power\,-\,law $f(R)$
theories since one should have the possibility of explaining
microlensing events without any dark matter halo (which is not
needed in the $f(R)$ theory we are considering). As a test, one
could try to fit the Milky Way and LMC rotation curves using the
modified gravitational potential and the mass distributions of the
above quoted models. A successful result should be a serious
evidence against dark matter halo in our Galaxy.

The pointlike lens is only the first and essential step towards
the formulation of a theory of gravitational lensing in a
power\,-\,law $f(R)$ theory. As a second step, we have to
investigate the case of an extended lens such as a galaxy. This
will offer us the possibility to get strong constraints on the
theory parameters $(\beta, \log{\tilde{v}_1})$ by fitting to the
images configuration in multiply imaged quasars. This will also
represent a critical test of the $f(R)$ theory since it is
expected that $\beta$ (and hence $n$) is the same for all the
galaxy lenses considered. With more than 90 systems (see, e.g.,
the CASTLES survey website \cite{castles}), this test will be
statistically meaningful and will be presented in a forthcoming
paper.

As a final remark, we briefly comment upon the possibility of
evaluating the corrected deflection angle for other classes of
$f(R)$ theories. In principle, one should only insert the
corresponding expression for the modified gravitational potential
$\Phi$ in the general equation (\ref{eq: alphacent}) for the
deflection angle. Given the fourth order degree of the field
equations, it is likely that the solution for $\Phi$ may still be
approximated (at least, within a limited range) by Eq.(\ref{eq:
phi}) with ($\beta, \log{\tilde{v}_1}$) related to the $f(R)$
parameters. It is therefore likely that the results discussed here
for $R^n$ gravity may be qualitatively extended to a more general
class of higher order theories of gravity.

\end{document}